\documentclass[11pt]{article}

\usepackage{amsmath}
\usepackage{amsthm}
\usepackage{amssymb}
\usepackage{amscd}
\usepackage{color}

\theoremstyle{definition}
\newtheorem{theorem}{Theorem}[section]
\newtheorem{prop}[theorem]{Proposition}
\newtheorem{lemma}[theorem]{Lemma}

\newtheorem{definition}[theorem]{Definition}
\newtheorem{example}[theorem]{Example}
\newtheorem{remark}[theorem]{Remark}

\numberwithin{equation}{section}
\newcommand\bea{\begin{eqnarray}}
\newcommand\ena{\end{eqnarray}}
\newcommand\non{\nonumber}
\newcommand\ord{\operatorname{ord}}

\allowdisplaybreaks

\title
{One Step Degeneration of Trigonal Curves and Mixing of Solitons and Quasi-Periodic Solutions
of the KP Equation
  \\
}

\author{
Atsushi Nakayashiki
\footnote{
Department of Mathematics, Tsuda University,
Kodaira, Tokyo 187-8577, Japan,
{\tt atsushi@tsuda.ac.jp}
}
}

\date{
}

\begin{document}

\maketitle

\begin{flushright}
\centerline{\it To the memory of Victor Enolski}
\end{flushright}

\begin{abstract}
We consider certain degenerations of trigonal curves and hyperelliptic curves, which we call 
one step degeneration. We compute the limits of corresponding quasi-periodic solutions 
using the Sato Grassmannian. The mixing of solitons and quasi-periodic 
solutions is clearly visible in the obtained solutions.
\end{abstract}

\section{Introduction}
The aim of this paper is  to compute explicitly the limits of quasi-periodic solutions of the KP equation
according to certain degenerations of trigonal and hyperelliptic curves, which we call 
oen step degeneration.

The KP equation is the $2+1$ dimensional equation given by 
\bea
&&
3u_{t_2 t_2}+(-4u_{t_3}+6uu_{t_1}+u_{t_1t_1t_1})_{t_1}=0,
\label{i-1}
\ena
where $(t_1,t_2)$ and $t_3$ are space and time variables respectively.
It can be rewritten in the Hirota bilinear form:
\bea
&&
(D_{t_1}^4-4D_{t_2}D_{t_3}+3D_{t_2}^2)\tau\cdot\tau=0,
\label{i-2}
\ena
where $D_{t_i}$'s are Hirota derivatives defined by
\bea
&&
f(t+s)g(t-s)=\sum_{n=0}^\infty D_{t}^{n} f\cdot g\frac{s^n}{n!}.
\non
\ena
For a solution $\tau$ of (\ref{i-2}) $u=2(\log \tau)_{t_1t_1}$
gives a solution of (\ref{i-1}). 
The KP hierarchy is the infinite system of differential equations which contains the KP equation (\ref{i-2})
as its first member \cite{DJKM}.  It is given by 
\bea
&&
\int\tau(t-s-[z^{-1}])\tau(t+s+[z^{-1}]){\rm e}^{-2\sum_{j=1}^\infty s_j z^j} \frac{dz}{2\pi i}=0,
\label{i-3}
\ena
where $t=(t_1,t_2,...)$, $s=(s_1,s_2,...)$, $[z^{-1}]=[z^{-1},z^{-2}/2,z^{-3}/3,...]$
 and the integral signifies taking the coefficient of $z^{-1}$ in the series expansion of the integrand.
Expanding (\ref{i-3}) by $s$ we get differential equations for $\tau(t)$ 
in the Hirota bilinear form. A solution $\tau(t)$ is sometimes called a tau function.
The introduction of the infinitely many variables is indispensable to the Sato theory which 
we use in this paper.

The KP hierarchy has a variety of solutions. Among them soliton solutions and algebro-geometric 
solutions are relevant to us. Soliton solutions are the solutions expressed by exponential functions 
given as follows (see \cite{Kodama2017} for example). Take positive integers $N<M$, non-zero
parameters $\kappa_j$, $1\leq j\leq M$ and 
an $N\times M$-matrix $A=(a_{i,j})$. Then soliton solution is given by
\bea
&&
\tau(t)=\sum_{I=(i_1<\cdots<i_N)}\Delta_IA_I{\rm e}^{\eta(\kappa_{i_1})+\cdots+\eta(\kappa_{i_M})},
\label{i-4}
\\
&&
\Delta_I=\prod_{p<q}(\kappa_{i_q}-\kappa_{i_p}),
\quad
A_I=\det(a_{p,i_q})_{1\leq p,q\leq N},
\quad
\eta(\kappa)=\sum_{i=1}^\infty t_i \kappa^i.
\non
\ena
Recently it was discovered that the shapes of soliton solutions form various web patterns and that 
they are related with the geometry of Grassmann manifolds, cluster algebras (see \cite{Kodama2017} and 
references therein).

Quasi-periodic solutions, whcih is also called algebro-geometric solutions, constitute a class of solutions expressed by theta functions of algebraic curves
with positive genus. Periodic solutions are contained in this class. 
 Soliton solutions can be considered as the limits of quasi-periodic solutions when periods go to infinity . 
In terms of curves soliton solutions are the genus zero limits of quasi-periodic solutions.
  Our original motivation of the research
was to take these limits and compare the structure of the quasi-periodic solutions  and that of solitons
described in \cite{Kodama2017}. However in the course of study \cite{BEN} we come to the recognition that the limits to positive genus solutions are more fundamental.
Anyhow the difficulty here is that to take a limit of a theta function or, in other words, to take a limit of the period matrix of an algebraic curve, is not very easy.

In \cite{N2018-1}\cite{N2018-2}\cite{BEN} we have demonstrated that the Sato Grassmannian (UGM) approach to this kind of problem is very effective. The reason, roughly speaking, is explained as follows. 
There is a one to 
one correspondence between points of UGM and solutions of the KP-hierarchy up to constants. Using 
UGM an algebro-geometric solution can be described as a series whose coefficients are constructed from 
some rational functions on an algebraic curve. In this way
the difficult problem on taking limits of period matrices reduces to much easier problem on taking limits 
of rational functions. In this paper we develop the UGM approach further.

We consider the following degeneration of algebraic curves, which  we call one step degeneration, given by
\bea
&&
y^m=\prod_{j=1}^{mn+1}(x-\alpha_j)
\hskip5mm
\longrightarrow
y^m=(x-\alpha)^m\prod_{j=1}^{m(n-1)+1}(x-\alpha_j),
\label{i-5}
\ena
for $m=2,3$. Fix $m$ and denote by $C_n$ the non-singular curve before taking the limit. We define
some canonical tau function $\tau_{n,0}(t)$ (see (\ref{tau0}) ) corresponding to the curve $C_n$. Then we express the limit of $\tau_{n,0}(t)$ in terms of $\tau_{n-1,0}(t)$ with the variable $t$ being appropriately shifted.
Then a solitonic structure can be seen clearly in the degeneration of  the algebro-geometric solution 
$\tau_{n,0}(t)$. This is another crucial idea in this paper.

The results are as follows.
For $m=2$, that is, the case of a hyperelliptic curve, we have (Theorem \ref{h-main-theorem}),
\bea
&&
\lim \tau_{n,0}(t)
=C{\rm e}^{-2\sum_{l=1}^\infty \alpha^l t_{2l}}
\non
\\
&&
\times\Bigl(
{\rm e}^{\eta(\alpha^{1/2})} \tau_{n-1,0}(t-[\alpha^{-1/2}])
+
(-1)^n{\rm e}^{\eta(-\alpha^{1/2})} \tau_{n-1,0}(t-[-\alpha^{-1/2}])
\Bigr),
\label{i-6}
\ena
for some constant $C$. It is observed that the soliton factors
$\displaystyle {\rm e}^{\eta(\pm \alpha^{1/2})}$ pop out from $\tau_{n,0}(t)$.
Then the solution (\ref{i-6}) looks the mixture of solitons and quasi-periodic solutions.
Using the formula repeatedly and noting that $\tau_{0,0}(t)=1$ if $\alpha_1=0$
 we get well known soliton solutions of the KdV equation.

For $m=3$ we have (Theorem \ref{main-theorem})
\bea
&&
\lim \tau_{n,0}(t)
\non
\\
&&
=
{\rm e}^{-6\sum_{l=1}^\infty\alpha^l t_{3l}}
\sum_{0\leq i<j\leq 2, 0\leq k\leq 2}
\frac{\partial}{\partial \beta}\Bigl(
\tilde{C}_{i,j,k}(\alpha,\beta){\rm e}^{\eta(z_i(\alpha)^{-1})+\eta(z_j(\alpha)^{-1})+\eta(z_k(\beta)^{-1})}
\non
\\
&&
\times
\tau_{n-1,0}(t-[z_i(\alpha)]-[z_j(\alpha)]-[z_k(\beta)])
\left.\Bigr)\right|_{\beta=\alpha},
\non
\\
&&
z_i(\alpha)=\omega^{-i}\alpha^{-1/3},
\quad
\omega={\rm e}^{2\pi i/3},
\label{i-7}
\ena
for some constants $\tilde{C}_{i,j,k}(\alpha,\beta)$. 
A new feature in this case is the appearence of the derivative 
with respect to the parameter $\beta$. This corresponds to the fact that the limit of $\tau_{n,0}$
to genus zero curve in this case is not a soliton but a generalized soliton \cite{N2018-1}.
The constants $\tilde{C}_{i,j,k}(\alpha,\beta)$ should be expressed by some derivatives of the 
sigma function. The explicit formulas for them are important for the further analysis of the solutions.

We remark that the formula of the forms (\ref{i-6}), (\ref{i-7}) can be generalized for $m\geq 4$ in (\ref{i-5}).
They should be treated  in a subsequent papers. A Generalization of the results in this paper to 
other class of curves such as that treated in \cite{AN} is also interesting.

The paper is organized as follows. In section 2 we first review the theory of the Sato Grassmannian (UGM).
Then we explain how to embedd the space of functions on an algebraic curve to UGM. Next we apply the general theory to our concrete examples and define the frame $\tilde{\xi}_n$ of a point of UGM corresponding to the space of regular rational functions on $C_n\backslash\{\infty\}$. Then we study the degeneration of 
$\tilde{\xi}_n$
and define the frame $\xi_n$ as a gauge transformation of $\tilde{\xi}_n$. In order to express $\xi_n$ by 
an object associated with the curve $C_{n-1}$ we study the frame associated with the space of rational functions on $C_{n-1}\backslash\{\infty\}$ which are singular at three points. Decomposing some 
rational functions we derive the degeneration formula of the tau function $\tau(t;\tilde{\xi}_n)$ 
corresponding to $\tilde{\xi}_n$ in terms of some tau functions associated with the curve $C_{n-1}$
 in the final subsection of section 2.
In section 3 we first review the sigma function of a so called $(N,M)$ curve. Then we recall the sigma 
function expression of  $\tau(t;\tilde{\xi}_n)$. Next we express the tau function corresponding to the space of functions with additional singularities as a shift of  $\tau(t;\tilde{\xi}_n)$. By subsituting these formulas to 
the degeneration formula derived in section 2 we express the limit of  $\tau(t;\tilde{\xi}_n)$ in terms 
of the shift of $\tau(t;\tilde{\xi}_{n-1})$. 
In section 4 we derive a similar degeneration formula for hyperelliptic curves based on the results of \cite{BEN}.

\section{Sato Grassmannian and $\tau$-function}
In this section we beriefly recall the definition and basic properties of the Sato Grassmannian,
\subsection{Sato Grassmannian}

Let $V={\mathbb C}((z))$ be the vector space of Laurent series in the variable $z$ and
$V_\phi={\mathbb C}[z^{-1}]$, $V_0=z{\mathbb C}[[z]]$ two subspaces of $V$.
Then $V$ is isomorphic to $V_\phi\oplus V_0$. Let $\pi:V\longrightarrow V_\phi$ be the projection
map. Then the Sato Grassmannian UGM is defined as the set of subspaces $U$ of  $V$ such that 
the restriction $\pi|_U$ has the finite dimensional kernel and cokernel whose dimensions coincide.

To an element $\sum a_n z^n\in V$ we associate the infinite column vector $(a_n)_{n\in {\mathbb Z}}$.
Then a frame of a point $U$ of UGM is expressed by an ${\mathbb Z}\times {\mathbb N}_{\leq0}$ matrix 
$\xi=(\xi_{i,j})_{i\in {\mathbb Z},j \in {\mathbb N}_{\leq0}}$, where columns, and therefore a basis 
of $U$, are labeled by the set of non-positive integers ${\mathbb N}_{\leq 0}$. 
A frame $\xi$ is written in the form
\bea
&&
\xi=
\left(
\begin{array}{ccc}
\quad&\vdots&\vdots\\
\cdots&\xi_{-1,-1}&\xi_{-1,0}\\
\cdots&\xi_{0,-1}&\xi_{0,0}\\
---&---&---\\
\cdots&\xi_{1,-1}&\xi_{1,0}\\
\cdots&\xi_{2,-1}&\xi_{2,0}\\
\quad&\vdots&\vdots\\
\end{array}
\right).
\ena
It is always possible to take a frame satisfying the following condition,
there exists a negative integer $l$ such that
\bea
&&
\xi_{i,j}=\left\{
\begin{array}{cl}
1&\text{ if $j<l$ and $i=j$ }\\
0&\text{ if ($j<l$ and $i<j$) or ($j\geq l$ and $i<l$)}.
\end{array}
\right.
\label{frame-cond}
\ena
In the sequel we always take a frame which satisfies this condition, although it is not unique.

A Maya diagram $M=(m_j)_{j=0}^\infty$ is a sequence of decreasing intergers  such that $m_j=-j$ for all 
sufficiently large $j$. For a Maya diagram $M=(m_j)_{j=0}^\infty$ the corresponding partition is defined by 
$\lambda(M)=(j+m_j)_{j=0}^\infty$. By this correspondence the set of Maya diagrams and the set 
of partitions bijectively correspond to each other. 

For a frame $\xi$ and a Maya diagram $M=(m_j)_{j=0}^\infty$ define the Pl\"ucker 
coordinate by
\bea
&&
\xi_M=\det(\xi_{m_i,j})_{-i,j\leq0}
\non
\ena
Due to the condition (\ref{frame-cond}) and the condition of the Maya diagram $M$ this infinite determinant can be computed as the finite determinant $\det(\xi_{m_i,j})_{k\leq -i,j\leq0}$ for sufficiently small $k$.

Define the elementary Schur function $p_n(t)$ by
\bea
&&
{\rm e}^{\sum_{n=1}^\infty t_n\kappa^n}=\sum_{n=0}^\infty p_n(t) \kappa^n.
\non
\ena
The Schur function \cite{Mac1995} corresponding to a partition $\lambda=(\lambda_1,...,\lambda_l)$ is defined by
\bea
&&
s_\lambda(t)=\det(p_{\lambda_i-i+j}(t))_{1\leq i,j\leq l}.
\non
\ena
Assign the weight $j$ to the variable $t_j$. Then it is known that $s_\lambda(t)$ is homogeneous 
of weight $|\lambda|=\lambda_1+\cdots+\lambda_l$.
To a point $U$ of UGM take a frame $\xi$ and define the tau function by
\bea
&&
\tau(t;\xi)=\sum_M \xi_M s_{\lambda(M)}(t).
\label{tau}
\ena
If we change the frame $\xi$ $\tau(t;\xi)$ is multiplied by a constant. 
We call $\tau(t;\xi)$, for any frame $\xi$ of $U$, a tau function corresponding to $U$.
So tau functions of a point of UGM differ by constant multiples to each other.

Then 
\begin{theorem}\cite{SS}\label{Sato-Sato}
The tau function $\tau(t;\xi)$ is a solution of the KP-hierarchy. Conversely for a 
formal power series solution $\tau(t)$ of the KP-hierarchy there exists a point $U$ of UGM 
such that $\tau(t)$ coincides with a tau function of $U$.
\end{theorem}

The point $U$ of UGM corresponding to a solution $\tau(t)$ in Theorem \ref{Sato-Sato} is 
given as follows \cite{SS,SN,KNTY,N2010-2}.

Let $\Psi^\ast(t;z)$ be the adjoint wave function \cite{DJKM} corresponding to $\tau(t)$ 
which is defined by
\bea
&&
\Psi^\ast(t;z)=\frac{\tau(t+[z])}{\tau(t)}{\rm e}^{-\sum_{i=1}^\infty t_iz^{-i}}.
\label{wave}
\ena
Define $\Psi_i^\ast(z)$ by the following expansion
\bea
\left(\tau(t)\Psi^\ast(t;z)\right)\vert_{t=(x,0,0,0,...)}
&=&
\tau((x,0,0,0,...)+[z]){\rm e}^{-x z^{-1}}
\non
\\
&=&
\sum_{i=0}^\infty \Psi_i^\ast(z) x^i.
\label{wave-expand}
\ena
Then
\bea
&&
U=\sum_{i=0}^\infty {\mathbb C}\Psi_i^\ast(z).
\label{U-tau}
\ena

By this correspondence between points of UGM and tau functions the following property follows.
Let $U$ be a point of UGM, $\tau(t)$ be a tau function corresponding to $U$ and 
$f(z)={\rm e}^{\sum_{i=1}^\infty a_i \frac{z^i}{i}}$ be an invertible formal power series.
Then $f(z)U$ belongs to UGM and the corresponding tau function is given by
\bea
&&
{\rm e}^{\sum_{i=1}^\infty a_i t_i}\tau(t).
\label{gauge-trf}
\ena
It is sometimes called the gauge transformation of $\tau(t)$.

\subsection{Embedding of algebro-geometric data to UGM}
In this section we recall the construction of points of UGM from alebraic curves (see \cite{Mul1994},\cite{N2018-2} for more details).

Let $C$ be a compact Riemann surface of genus $g$, $p_\infty$ a point on it, $z$ a local coordinate around 
$p_\infty$.
For $m\geq 0$ and points $p_i$, $1\leq i\leq m$, on $C$, such that $p_j\neq\infty$ for any $j$, we denote by 
\bea
&&
H^0(C,{\cal O}(\sum_{j=1}^m p_j+\ast p_\infty))
\label{m-points}
\ena
 the vector space of meromorphic functions on $C$
which have a pole at each $p_j$ of order at most $1$ and have  a pole at $p_\infty$ of any order.
By expanding functions in the local coordinate $z$ we can consider
$H^0(C,{\cal O}(\sum_{j=1}^m p_j+\ast p_\infty))$ as a subspace of $V={\mathbb C}((z))$.
Then

\begin{prop}{\rm  \cite{N2018-2}\cite{Mul1994}}\label{UGM-1}
The subspace $z^{g-m}H^0(C,{\cal O}(\sum_{j=1}^m p_j+\ast p_\infty))$ belongs to UGM.
\end{prop}

\begin{remark}
This Proposition was proved in \cite{N2018-2} from the general results \cite{Mul1994}, for $m\leq g$. 
But the case $m>g$ can be proved in the same way.
\end{remark}

\subsection{Tau function corresponding to zero point space}
For $n\geq 1$ and  mutually distinct complex numbers $\{\alpha_i\}_{i=1}^{3n}$ consider the compact Riemann surface $C_n$ corresponding to  the algebraic curve  defined by the equation
\bea
&& y^3=\prod_{j=1}^{3n+1}(x-\alpha_j).
\label{C-1}
\ena
The genus of $C_n$ 
is $g=3n$ and there is a unique point on $C_n$ over $x=\infty$ which we denote by $\infty$.

Consider the space $H^0(C_n,{\cal O}(\ast \infty))$ which corresponds to $m=0$ in (\ref{m-points}).
It is the space of meromorphic functions on $C$ which are regular on $C_n\backslash\{\infty\}$. 
It can be easily proved that it coicides with the vector space ${\mathbb C}[x,y]$ of polynomials 
in $x,y$.
A basis of this vector space is given by
\bea
&&
x^i , 
\hskip5mm
x^iy, 
\hskip5mm
x^iy^2
\quad i\geq 0.
\label{B-1}
\ena
We take the local coordinate $z$ around $\infty$ such that 
\bea
&&
x=z^{-3},
\hskip5mm
y=z^{-(3n+1)}F_n(z), 
\hskip5mm
F_n(z)=\left(\prod_{j=1}^{3n+1}(1-\alpha_j z^3)\right)^{1/3}.
\label{coordinate1}
\ena
In the following we denote by $z$ this local coordinate unless otherwise stated.
The function $F_n(z)$ is considered as a power series in $z$ by the Taylor expansion at $z=0$.

By Proposition \ref{UGM-1} $z^gH^0(C_n,{\cal O}(\ast \infty))$ determines a point of UGM.
Writing  (\ref{B-1}) in terms of $z$ and multiplying them by $z^g$ we get a basis of  it,

\bea
&&
z^{3n-3i} , 
\hskip5mm
z^{-1-3i}F_n(z), 
\hskip5mm
z^{-3n-2-3i}F_n(z)^2\quad
\quad i\geq 0.
\label{B-2}
\ena
We define the frame $\tilde{\xi}_n$ from this basis as follows.

For an element $v(z)=\sum_{n\leq i}a_i z^i$, $a_n\neq 0$, define the order of $v(z)$ to be $-n$ and write 
$\ord v(z)=-n$.

\begin{definition}
Label the elements of  (\ref{B-2}) by $\tilde{v}_i$, $i\leq 0$, in such a way that 
$\ord \tilde{v}_0<\ord \tilde{v}_{-1}<\ord \tilde{v}_{-2}<\cdots$ and define 
the frame $\tilde{\xi}_n$ of $z^g H^0(C_n, {\cal O}(\ast \infty))$ by
\bea
&&
\tilde{\xi}_n=(\ldots,\tilde{v}_{-2},\tilde{v}_{-1}, \tilde{v}_0).
\label{frame-txi}
\ena
\end{definition}
By the construction of $\tilde{\xi}_n$ the tau function $\tau(t;\tilde{\xi}_n)$ 
has the following expansion (see \cite{N2010-2})
\bea
&&
\tau(t;\tilde{\xi}_n)=s_{\lambda^{(n)}}(t)+\rm{h.w.t},
\label{tau-expansion-1}
\ena
where h.w.t means the higher weight terms,  $\lambda^{(n)}$ is the partition determined from the gap sequence $w_1<\cdots<w_g$ at $\infty$ of $C_n$ and is given by
\bea
&&
\lambda^{(n)}=(w_g-(g-1),...,w_2-1,w_1).
\non
\ena

\begin{example} 
$\lambda^{(1)}=(3,1,1)$,
$\lambda^{(2)}=(6,4,2,2,1,1)$,
$\lambda^{(3)}=(9,7,5,3,3,2,2,1,1)$
\end{example}

\subsection{Degeneration}

Let us take a complex number $\alpha$ which is different from $\alpha_i$, $1\leq i\leq 3n-2$
and consider the limit 
\bea
&&
\alpha_{3n+1}, \alpha_{3n}, \alpha_{3n-1}\to \alpha,
\label{limit}
\ena
which means that the curve $C_n$ degenerates to
\bea
&&
y^3=(x-\alpha)^3\prod_{j=1}^{3n-2}(x-\alpha_j).
\label{C-2-1}
\ena
which we call one step degeneration of $C_n$.

In the limit 
\bea
&&
F_n(z)\hskip2mm \longrightarrow \hskip2mm(1-\alpha z^3)F_{n-1}(z),
\non
\ena
and the basis  (\ref{B-2}) tends to 
\bea
&&
z^{3n-3i} , 
\hskip1mm
z^{-1-3i}(1-\alpha z^3)F_{n-1}(z), 
\hskip1mm
z^{-3n-2-3i}(1-\alpha z^3)^2F_{n-1}(z)^2,
\, i\geq 0.
\label{B-3}
\ena
Let $W_n$ be the point of UGM generated by this basis.
Multiply (\ref{B-3}) by $(1-\alpha z^3)^{-2}$ we have
\bea
&&
\frac{z^{3n-3i}}{(1-\alpha z^3)^2} , 
\hskip5mm
\frac{z^{-1-3i}}{(1-\alpha z^3)}F_{n-1}(z), 
\hskip5mm
z^{-3n-2-3i}F_{n-1}(z)^2\quad
\quad i\geq 0.
\label{B-4}
\ena

By taking linear combinations we have

\begin{lemma}
The following set of elements gives a basis of $(1-\alpha z^3)^{-2}W_n$.
\bea
&&
z^{3n-6-3i},
\hskip3mm
z^{-4-3i}F_{n-1}(z),
\hskip3mm
z^{-3n-2-3i}F_{n-1}(z)^2,
\hskip3mm
i\geq 0,
\non
\\
&&
\frac{z^{3n}}{(1-\alpha z^3)^2},
\hskip3mm
\frac{z^{3n-3}}{1-\alpha z^3},
\hskip3mm
\frac{z^{-1}}{1-\alpha z^3}F_{n-1}(z).
\label{B-5}
\ena
\end{lemma}

We arrange the basis elements  of this lemma according as their orders and 
define the frame $\xi_n$ as follows.

\begin{definition}
Define the frame $\xi_n$ of $W_n$ by
\bea
&&
\xi_n=(\ldots,v_{-2},v_{-1},v_0),
\non
\ena
with 
\bea
v_0&=&\frac{z^{3n}}{(1-\alpha z^3)^2},
\non
\\
v_{-1}&=&\frac{z^{3n-3}}{1-\alpha z^3},
\non
\\
v_{-(2+i)}&=&z^{3n-6-3i}, \hskip18mm 0\leq i\leq n-2,
\non
\\
v_{-(n+1)}&=&\frac{z^{-1}}{1-\alpha z^3}F_{n-1}(z),
\non
\\
v_{-(n+2+2i)}&=&z^{-3-3i}, \hskip23mm 0\leq i\leq n-1,
\non
\\
v_{-(n+3+2i)}&=&z^{-4-3i}F_{n-1}(z),  \hskip10mm 0\leq i\leq n-1,
\non
\\
v_{-(3n+2+3i)}&=&z^{-3n-2-3i}F_{n-1}(z)^2,\hskip5mm  i\geq 0,
\non
\\
v_{-(3n+3+3i)}&=&z^{-3n-3-3i}, \hskip20mm i\geq 0,
\non
\\
v_{-(3n+4+3i)}&=&z^{-3n-4-3i}F_{n-1}(z),\hskip8mm  i\geq 0.
\non
\ena
\end{definition}

Since we have the expansion
\bea
&&
\log (1-\alpha z^3)^{-2}=6\sum_{l=1}^\infty\alpha^l\frac{z^{3l}}{3l},
\non
\ena
 the following relation holds by  (\ref{gauge-trf}), 
\bea
&&
 \tau(t;{\xi}_n)={\rm e}^{6\sum_{l=1}^\infty \alpha^l t_{3l}}\lim \tau(t;\tilde{\xi}_n),
\label{xi-txi}
\ena
where the $\lim$ signifies taking the limit (\ref{limit}).

\subsection{Three point insertion}
Consider the curve $C_{n-1}$ defined by (\ref{C-1}) where $n$ is replaced by $n-1$.
The genus of $C_{n-1}$ is $g'=3n-3=g-3$.
Let 
\bea
&&
Q_j=(c_j,Y_j),\hskip3mm j=0,1,2,
\label{Q-j}
\ena
be points on $C_{n-1}$. We assume $c_j\neq \alpha_i$ for any $i,j$. 
Define $\varphi_j$ by
\bea
&&
\varphi_j=\frac{y^2+Y_jy+Y_j^2}{x-c_j}.
\non
\ena
The pole divisor of this function is $Q_j+(2g'-1)\infty$.
Consider the space 
$H^0(C_{n-1},{\cal O}(Q_0+Q_1+Q_2+\ast\infty))$.
A basis of it is given by
\bea
&&
x^i,
\hskip3mm
x^iy,
\hskip3mm
x^iy^2,
\hskip3mm
\varphi_j,
\hskip3mm
i\geq0,\quad j=0,1,2.
\non
\ena
Write this basis in terms of the local coordinate $z$ and multiply it by $z^{g'-3}$ we have
\bea
&&
z^{3n-6-3i},
\hskip1mm
z^{-4-3i}F_{n-1}(z),
\hskip1mm
z^{-3n-2-3i}F_{n-1}(z)^2,
\,
z^{3n-6}\varphi_j,
\,
i\geq0,\, j=0,1,2.
\label{basis-3points-1}
\ena
By Proposition \ref{UGM-1} $z^{g'-3}H^0(C_{n-1},{\cal O}(Q_0+Q_1+Q_2+\ast\infty))$ is a point of UGM 
and the set of functions (\ref{basis-3points-1}) is a basis of it.
Using this basis define the frame of  $z^{g'-3}H^0(C_{n-1},{\cal O}(Q_0+Q_1+Q_2+\ast\infty))$ by
\bea
&&
\xi_{n-1}(Q_0,Q_1,Q_2)=(\ldots,v_{-(n+3)},v_{-(n+2)},v_{-n},...,v_{-2},z^{3n-6}\psi_0,z^{3n-6}\psi_1,z^{3n-6}\psi_2),
\non
\ena
where $v_j$ is the same as that  in $\xi_n$.

\subsection{Degeneration formmula in algebraic form}
Corresponding to the parameter $\alpha$ in (\ref{limit}) let
$P_i(\alpha)=(\alpha,\omega^i y_0(\alpha))$, $i=0,1,2$ be points on $C_{n-1}$, where $\omega={\rm e}^{2\pi i/3}$.
Take $Q_j=P_j(\alpha)$ in (\ref{Q-j}) and denote the function $\varphi_j$ by $\varphi_j(\alpha)$. 
Then
\bea
&&
\varphi_j(\alpha)=\frac{y^2+(\omega^j y_0(\alpha))y+(\omega^j y_0(\alpha))^2}{x-\alpha}.
\non
\ena

\begin{lemma}
For $0\leq i\leq 2$ we have
\bea
&&
\frac{y^i}{x-\alpha}=\frac{1}{3y_0(\alpha)^{2-i}}\sum_{j=0}^2 \omega^{(i+1)j}\varphi_j(\alpha)
\non
\ena
\end{lemma}

The lemma can be verified by direct computation.
From these relations we have
\bea
v_{-1}
&=&
\frac{z^{3n-3}}{1-\alpha z^3}
=
\frac{1}{3y_0(\alpha)^2}\sum_{i=0}^2\omega^i z^{3n-6}\varphi_i(\alpha)
\label{decomposition-1}
\\
v_{-(n+1)}&=&
\frac{z^{-1}F_{n-1}(z)}{1-\alpha z^3}
=
\frac{1}{3y_0(\alpha)}\sum_{i=0}^2\omega^{2i} z^{3n-6}\varphi_i(\alpha)
\label{decomposition-2}
\\
v_0&=&
\frac{z^{3n}}{(1-\alpha z^3)^2}
=
\frac{\partial}{\partial \beta}\left.\left(
\frac{1}{3y_0(\beta)^2}\sum_{i=0}^2\omega^i z^{3n-6}\varphi_i(\beta)\right)\right|_{\beta=\alpha}.
\label{decomposition-3}
\ena
 The third equation is obtained by differentiating the first equation in $\alpha$.
 
Let $\lambda$ be a partition and consider the Pl\"ucker coordinate of $(\xi_{n})_\lambda$.
Substitute the above expression to the definition of $(\xi_{n})_\lambda$ of $\xi_n$.
Then Equations (\ref{decomposition-1})-(\ref{decomposition-3}) mean that each of the column vectors
 of $\xi_n$ 
corresponding  to $v_0$, $v_{-1}$, $v_{-(n+1)}$ is a sum of vectors. So we have 

\bea
(\xi_{n})_\lambda&=&\frac{(-1)^n}{27y_{n-1,0}(\alpha)^5}
\sum_{0\leq i<j\leq 2, 0\leq k\leq 2}\omega^{i+k+2j}(1-\omega^{i-j})
\non
\\
&&\qquad
\times
\frac{\partial}{\partial \beta}\left(\xi_{n-1}(P_{i}(\alpha),P_{j}(\alpha),P_{k}(\beta))_\lambda\right)\vert_{\beta=\alpha}.
\non
\ena
Multiplying this equation by $s_\lambda(x)$ and summing up in $\lambda$ we get

\bea
\tau(t;\xi_{n})&=&\frac{(-1)^n}{27y_{n-1,0}(\alpha)^5}
\sum_{0\leq i<j\leq 2, 0\leq k\leq 2}\omega^{i+k+2j}(1-\omega^{i-j})
\non
\\
&&\qquad
\times
\frac{\partial}{\partial \beta}\left(\tau(t;\xi_{n-1}(P_{n-1,i}(\alpha),P_{n-1,j}(\alpha),P_{n-1,k}(\beta)))\right)\vert_{\beta=\alpha}.
\non
\ena

Finally using (\ref{xi-txi}) we obtain 

\begin{theorem}\label{relation-1}
Consider the limit (\ref{limit}).
Then the limit of the tau function of the frame $\tilde{\xi}_n$ defined by (\ref{frame-txi})  is given by the following formula:
\bea
\lim \tau(t;\tilde{\xi}_{n})&=&\frac{(-1)^n}{27y_{n-1,0}(\alpha)^5}{\rm e}^{-6\sum_{l=1}^\infty\alpha^l t_{3l}}
\sum_{0\leq i<j\leq 2, 0\leq k\leq 2}\omega^{i+k+2j}(1-\omega^{i-j})
\non
\\
&&\qquad
\times
\frac{\partial}{\partial \beta}\left.\Bigl(\tau\left(t;\xi_{n-1}(P_{i}(\alpha),P_{j}(\alpha),P_{k}(\beta))\right)\Bigr)\right|_{\beta=\alpha}.
\non
\ena
\end{theorem}

\begin{remark} The new feature of the trigonal case compared with the hyperelliptic case studied 
in \cite{BEN} (see Theorem \ref{h-theorem1} ) is the existence of a derivative in the parameter $\beta$. 
In \cite{N2018-1} the degeneration to genus zero curve in the trigonal case was directly studied.
The obtained solutions are not solitons but generalized solitons. The appearence of the derivative
 corresponds to this phenomenon.
\end{remark}

\section{Analytic expression of tau functions}
In this section we derive the analytic expression of tau functions appeared in 
Theorem \ref{relation-1} in terms of the multivariate sigma function \cite{BEL1997}
\cite{BEL1999}\cite{BEL2012}\cite{N2010-1}\cite{N2010-2}. The fundamental idea 
behind constructing the expression is due to Krichever \cite{Kr1977}.

\subsection{The sigma function of an $(N,M)$ curve}
We consider the general $(N,M)$-curve \cite{BEL1999} defined by $f(x,y)=0$
with 
\bea
&&
f(x,y)=y^N-x^M-\sum_{Ni+Mj<NM}\lambda_{ij} x^i y^j,
\label{NMcurve}
\ena
where $N,M$ are relatively prime integers such that $1<N<M$. We assume that 
the curve is nonsingular. We denote the corresponding compact Riemann surface by $C$. 
Then the genus of $C$ is given by $g=1/2(N-1)(M-1)$. There is one point on $C$ over $x=\infty$ which 
is also denoted by $\infty$. Here we recall several necessary facts related with the curve $C$. See 
\cite{N2010-1,N2010-2} for details.

We assign the order $Ni+Mj$ to the monomial $x^iy^j$, $i,j\geq 0$, and define $f_i$, $i\geq 1$, to be the $i$-th monomial
in this order. For example $f_1=1$, $f_2=x$. Then the set of differentials
\bea
&&
du_i=-\frac{f_{g+1-i}dx}{f_y}, \quad 1\leq i\leq g
\non
\ena
constitutes a basis of holomorphic one forms. 
We choose an algebraic fundamental form $\widehat{\omega}(p_1,p_2)$ on $C\times C$ as in \cite{N2010-1}.
It has the decomposition of the form
\bea
&&
\widehat{\omega}(p_1,p_2)=d_{p_2}\Omega(p_1,p_2)+\sum_{i=1}^g du_i(p_1)dr_i(p_2),
\non
\ena
where $\Omega(p_1,p_2)$ is a certain meromorphic one form on $C\times C$ and $dr_i(p)$ is a certain differential of the second kind on $C$ 
with a pole only at $\infty$ (see \cite{N2010-1} for more precise form of $\widehat{\omega}$, $\Omega$, $dr_i$).
Taking a symplectic basis $\{\alpha_i,\beta_i\}_{i=1}^{g}$ of the homology group of $C$ we define 
the period matrices $\omega_k$, $\eta_k$, $k=1,2$, $\Pi$ by
\bea
&&
2\omega_1=\left(\int_{\alpha_j} du_i\right),
\qquad
2\omega_2=\left(\int_{\beta_j} du_i\right),
\non
\\
&&
-2\omega_1=\left(\int_{\alpha_j} dr_i\right),
\qquad
-2\omega_2=\left(\int_{\beta_j} dr_i\right),
\non
\ena
and $\Pi=\omega_1^{-1}\omega_2$.
Define  Riemann's theta function by
\bea
&&
\theta[\epsilon](z,\Pi)=\sum_{m\in {\mathbb Z}^g}{\rm e}^{\pi i {}^t (m+\epsilon')\Pi (m+\epsilon')
+2\pi i{}^t(m+\epsilon')(z+\epsilon'')},
\non
\ena
where $\epsilon={}^t(\epsilon',\epsilon'')\in {\mathbb R}^{2g}$, $\epsilon', \epsilon''\in {\mathbb R}^g$.
Let $\Pi \delta' +\delta''$, $\delta',\delta''\in (1/2){\mathbb Z}^g$, be a representative 
of Riemann's constant with respect to the choice of the base point $\infty$ and $\{\alpha_i,\beta_i\}_{i=1}^g$, and
$\delta={}^t(\delta',\delta'')\in (1/2){\mathbb Z}^{2g}$.

Let $(w_1,...,w_g)$, $w_1<\cdots<w_g$, be the gap sequence of  the curve $C$ at $\infty$ (see \cite{N2010-1},
\cite{FK} for example). Define the partition $\lambda^{(N,M)}$ by
\bea
&&
\lambda^{(N,M)}=(w_g-(g-1),...,w_2-1,w_1).
\non
\ena
By the definition $\lambda^{(n)}=\lambda^{(3,3n+1)}$.

\begin{definition}
The sigma function is defined by 
\bea
\sigma(u)&=&C{\rm e}^{\frac{1}{2}{}^t u \eta_1\omega_1^{-1} u}
\theta[-\delta]((2\omega_1)^{-1}u,\Pi),
\non
\\
u&=&{}^t(u_1,...,u_g)
\non
\ena
for some constant $C$. 
\end{definition}

Assign the weight $w_i$ to $u_i$.
Then the constant $C$ is specified by the condition that $\sigma(u)$ has the expansion 
of the form
\bea
&&
\sigma(u)=s_{\lambda^{(N,M)} }(t) \vert_{t_{w_i}=u_i}+{\rm h.w.t.}
\non
\ena
It is known that $C$ is explicitly expressed by some derivatives of the Riemann's theta 
function \cite{NY2012}\cite{N2016}.
The sigma function satisfies the following quasi-periodicity property: 
\bea
&&
\sigma(u+\sum_{i=1}^2 2\omega_i m_i)
=
\non
\\
&&
(-1)^{{}^t m_1 m_2+2({}^t\delta' m_1-{}^t\delta'' m_2)}
{\rm e}^{{}^t(\sum_{i=1}^2 2\eta_i m_i)(u+\sum_{i=1}^2 \omega_i m_i)} \sigma(u).
\label{quasi-periodicity}
\ena

\subsection{Sigma function expression of tau functions}
Here we derive sigma function expressions for the tau functions corresponding to 
the spaces in Proposition \ref{UGM-1} in the case of $(N,M)$ curves.

We take the local coordinate $z$ around $\infty$ such that
\bea
&&
x=z^{-N},
\qquad
y=z^{-M}(1+O(z)).
\non
\ena
Expand $du_i$, $\widehat{\omega}$ in $z$ as
\bea
du_i&=&\sum_{j=1}^\infty b_{i,j} z^{j-1},
\non
\\
\widehat{\omega}(p_1,p_2)&=&
\left(\frac{1}{(z_1-z_2)^2}+\sum_{i,j\geq 1}\widehat{q}_{i,j} z_1^{i-1} z_2^{j-1}\right) dz_1 dz_2,
\non
\ena
where $z_i=z(p_i)$. 
The differential $du_g$ has a zero of order $2g-2$ at $\infty$ and has the expansion 
of the form 
\bea
&&
du_g=z^{2g-2}(1+\sum_{j=2g}^\infty b_{g,j}z^{j-2g+1}) dz.
\non
\ena
Define $c_i$ by the expansion 
\bea
&&
\log\left(\sqrt{z^{2g-2}\frac{du_g}{dz}}\right)=
\sum_{i=1}^\infty c_i\frac{z^i}{i}.
\non
\ena
In \cite{N2010-2} there is a pisprint, $c_i z^i$ should be $c_i z^i/i$ as above.
Define $g\times {\mathbb N}$ matrix $B$ and the quadratic form $\widehat{q}$ by
\bea
&&
B=(b_{i,j})_{1\leq i\leq g, j\geq 1},
\qquad
\widehat{q}(t)=\sum_{i,j=1}^\infty \widehat{q}_{i,j} t_i t_j.
\non
\ena

The following theorem is proved in \cite{N2010-2}.

\begin{theorem}{\rm \cite{N2010-2}}\label{0-point-space}
A tau function correspodning to $z^gH^0(C,{\cal O}(\ast \infty))$ 
 is given by
\bea
&&
\tau_0(t):={\rm e}^{-\sum_{i=1}^\infty c_i t_i+\frac{1}{2}\widehat{q}(t)}\sigma(Bt).
\label{tau0}
\ena
It has the expansion of the form 
\bea
&&
\tau_{0}(t)=s_{\lambda^{(N,M)}}(t)+\rm{h.w.t.}
\label{tau0-expansion}
\ena
\end{theorem}

\begin{remark}\label{remark-tau0}
In \cite{N2010-2} it is proved that $\tau_0(t)$ defined by (\ref{tau0}) is a solution of the 
$N$-reduced KP-hierarchy \cite{DJKM}.
\end{remark}

More generally the tau function corresponding to the $m$-point space with $m\geq 1$ 
 given by Proposition \ref{UGM-1}
is described in terms of the shift of $\tau_0(t)$.

 \begin{theorem}\label{m-point-space}
Let $p_i$, $1\leq i\leq m$, be points on $C\backslash\{\infty\}$ and $z_i=z(p_i)$. 
A tau function corresponding to $z^{g-m}H^0(C,{\cal O}(\sum_{i=1}^m p_i+\ast \infty))$ is given by 
\bea
&&
\tau(t|p_1,...,p_m):={\rm e}^{\sum_{i=1}^\infty\eta(z_i^{-1})}\tau_0(t-\sum_{i=1}^m[z_i]),
\label{tau-m}
\ena
where $\eta(\kappa)=\sum_{i=1}^\infty t_i \kappa^i$, $[w]=[w,w^2/2,w^3/3,...]$.
\end{theorem}

By (\ref{gauge-trf}) and by that the KP-hierarchy is the system of autonomous equations, if  $\tau(t)$ is a solution of the KP-hierarchy, so is ${\rm e}^{\sum_{i=1}^\infty \gamma_i t_i}
\tau(t+\zeta)$ for any set of constants $\{\gamma_i\}$ and a constant vector $\zeta\in {\mathbb C}^g$. 
Therefore $\tau(t|p_1,...,p_m)$ is a solution of the KP-hierarchy.

Then the theorem is proved by calculating the adjoint wave function using (\ref{U-tau}).
To this end we need some notation.

Let $E(p_1,p_2)$ be the prime form \cite{Fay73} (see also \cite{KNTY}). 
Define $E(z_1,z_2)$, $E(q,p)$ with $z_i=z(p_i)$ and $q$ being a fixed point on $C$ by
\bea
&&
E(p_1,p_2)=\frac{E(z_1,z_2)}{\sqrt{d z_1}\sqrt{d z_2}},
\qquad
E(q,p)=\frac{E(z(q),z(p))}{\sqrt{d z(p)}}.
\non
\ena
Define $\tilde{E}(q,p)$ for $q$ fixed by
\bea
&&
\tilde{E}(q,p)=E(q,p)\sqrt{du_g(p)}{\rm e}^{\frac{1}{2}\int_q^p {}^t du (\eta_1\omega_1^{-1})\int_q^p du},
\non
\\
&&
du={}^t(du_1,\ldots,du_g).
\non
\ena
 In \cite{N2010-1} two variables $\tilde{E}(p_1,p_2)$ and one variable $\tilde{E}(\infty,p)$ were introduced
and studied.
It should be noticed that $\tilde{E}(q,p)$ is a multiplicative function of $p$ while $E(q,p)$ is a $-1/2$ form.
Similarly to the case of $\tilde{E}(\infty,p)$ in \cite{N2010-1} the following lemma can be proved.

\begin{lemma}\label{prime-function}
(i) The function $\tilde{E}(q,p)$ has the expansion in $z=z(p)$ near $\infty$ of the form
\bea
&&
\tilde{E}(q,p)=(z-z(q))z^{g-1}(1+O(z)).
\non
\ena
\vskip2mm
\noindent
(ii) Let $\gamma$ be an element of $\pi_1(C,\infty)$ and its Abelian image be 
$\sum_{i=1}^g(m_{1,i}\alpha_i+m_{2,i}\beta_i)$. Then
\bea
&&
\tilde{E}(q,\gamma(p))/\tilde{E}(q,p)
=
\non
\\
&&
(-1)^{{}^t m_1 m_2+2({}^t\delta' m_1-{}^t\delta'' m_2)}
{\rm e}^{{}^t(\sum_{i=1}^2 2\eta_i m_i)(\int_q^p du+\sum_{i=1}^2 \omega_i m_i)},
\label{quasi-periodicity-2}
\ena
where $m_i={}^t(m_{i,1},\ldots,m_{i,g})$.
\end{lemma}

By (i) of this lemma $\tilde{E}(\infty,p)$ has a zero of order $g$ at $\infty$.
\vskip2mm

Let $d\tilde{r}_i$ be the normalized differential of the second kind with a pole only at $\infty$, that is, it
satisfies 
\bea
&&
\int_{\alpha_j}d\tilde{r}_i=0, \quad 1\leq j\leq g,
\qquad
d\tilde{r}_i=d(z^{-i}+O(1)).
\non
\ena

Define
\bea
&&
d\hat{r}_i=d\tilde{r}_i+\sum_{j,k=1}^g b_{j,i}(\eta_1\omega_1^{-1})_{j,k} du_k.
\non
\ena
By the construction their periods can be computed as (Lemma 5 in \cite{N2010-2})
\bea
&&
\int_{\alpha_j} d\widehat{r}_i=\left({}^t(2\eta_1)B\right)_{j,i},
\hskip10mm
\int_{\beta_j} d\widehat{r}_i=\left({}^t(2\eta_2)B\right)_{j,i}.
\label{periods}
\ena
In Lemma 5 of \cite{N2010-2} there is a misprint: the right hand side is not the $(i,j)$ component 
but the $(j,i)$ component.

\noindent
{\it Proof of Theorem  \ref{m-point-space} }
\par

The adjoint wave function (\ref{wave}) corresponding to the tau function (\ref{tau-m}) is computed as
\bea
&&
\Psi^\ast(t,z)=C(z_1,...,z_m)z^{g-m}
\frac{\tilde{E}(\infty,p)^{m-1}\sigma\left(\int_\infty^p du-\sum_{i=1}^m \int_\infty^{p_i} du+Bt\right)}
{\prod_{i=1}^m \tilde{E}(p_i,p)\,\, \sigma\left(-\sum_{i=1}^m \int_\infty^{p_i} du+Bt\right)}
\non
\\
&&
\hskip20mm
\times{\rm e}^{-\sum_{i=1}^\infty t_i \int^p d\widehat{r}_i},
\non
\\
&&
C(z_1,...,z_m)=(-1)^m(\prod_{i=1}^mz_i)
{\rm e}^{\frac{1}{2}\sum_{i=1}^m \int_\infty^{p_i} {}^t du (\eta_1\omega_1^{-1})\int_\infty^{p_i} du}.
\non
\ena

By Lemma \ref{prime-function} and (\ref{periods}) we can check that $z^{-g+m}\Psi^\ast(t,z)$ is, as a function of $p\in C$,
$\pi_1(C,\infty)$ invariant.  Then the same is true for any expansion coefficient of $\Psi^\ast(t,z)$ in $t$.
Expansion coefficients in $t$ are regular except $p_i$, $1\leq i\leq m$, $\infty$ and have at most a simple pole at $p_i$.
Therefore the point $U$ of UGM corresponding to $\tau(t|p_1,...,p_m)$ is contained in $z^{g-m}H^0(C,{\cal O}(\sum_{i=1}^m p_i+\ast\infty))$. Since a strict inclusion relation is impossible for two points of UGM (Lemma 4.17 of \cite{BEN}), these two points of UGM coincide.
$\Box$

\subsection{Degeneration formula in analytic form}
In this section we apply the results in the previous section to the curves $C_n$, $C_{n-1}$ and associated tau
functions in Theorem \ref{relation-1}. So, in this section $\tau_{n,0}(t)$ denotes the function defined by (\ref{tau0}) for  the curve $C_n$.

\begin{lemma} We have
\bea
&&
\tau(t;\tilde{\xi}_n)=\tau_{n,0}(t).
\label{expression-1}
\ena
\end{lemma}
\vskip2mm
{\it Proof.} Since $\tilde{\xi}_n$ is a tau function corresponding to $z^gH^0(C_n,{\cal O}(\ast\infty))$, we have, 
by Theorem \ref{0-point-space}, 
\bea
&&
\tau(t;\tilde{\xi}_n)=C\tau_{n,0}(t),
\non
\ena
for some constant $C$. Comparing the expansions (\ref{tau-expansion-1}) and (\ref{tau0-expansion}) we have $C=1$. \qed

Next we consider tau functions appeared in the right hand side of the equation in Theorem \ref{relation-1}.
We need a point $(\alpha,y_0(\alpha))$ of $C_{n-1}$. To specify $y_0(\alpha)$ is equivalent to specify one value of $z$ such that  $z^{-3}=\alpha$, that is, $\alpha^{-1/3}$. In fact, if $z=\alpha^{-1/3}$ is given the value of $y_0(\alpha)$ is determined by (\ref{coordinate1})
as
\bea
&&
y_0(\alpha)=\alpha^{n-1}\alpha^{1/3}F_{n-1}(\alpha^{-1/3}).
\label{y0}
\ena
Since $P_i(\alpha)=(\alpha,\omega^iy_0(\alpha))$, we have
\bea
&&
z(P_{i}(\alpha))=\omega^{-i}\alpha^{-\frac{1}{3}}.
\label{z-coord-pi}
\ena
For simplicity we set
\bea
&&
z_i(\alpha)=\omega^{-i}\alpha^{-\frac{1}{3}}.
\label{zi-alpha}
\ena
Since,  in general $\xi_{n-1}(Q_0,Q_1,Q_2)$ is a frame of the point
\bea
&&
z^{g'-3}H^0(C_{n-1},{\cal O}(\sum_{i=0}^3 P_i+\ast\infty))\in {\rm UGM}
\non
\ena
 we have, by Theorem \ref{m-point-space}, 
\bea
&&
\tau(t;\xi_{n-1}(P_{i}(\alpha),P_{j}(\alpha),P_{k}(\beta)))
\non
\\
&&
=C_{i,j,k}(\alpha,\beta)
{\rm e}^{\eta(z_i(\alpha)^{-1})+\eta(z_j(\alpha)^{-1})+\eta(z_k(\beta)^{-1})}
\non
\\
&&
\times
\tau_{n-1,0}(t-[z_i(\alpha)]-[z_j(\alpha)]-[z_k(\beta)]),
\label{expression-2}
\ena
for some constsnt $C_{i,j,k}(\alpha,\beta)$.

\begin{remark} The explicit forms of the constants $C_{i,j,k}(\alpha,\beta)$ are not yet determined.
They should be calculated by comparing the Schur function expansions and are expected to be expressed by some derivatives of the sigma function. 
\end{remark}

Substituting (\ref{expression-1}), (\ref{expression-2}) into the relation in Theorem \ref{relation-1} we get

\begin{theorem}\label{main-theorem}
Let $\tau_{n,0}(t)$ be defined by the right hand side of  (\ref{tau0}) for the curve $C_n$ and $z_i(\alpha)$ defined by (\ref{zi-alpha}). Then, in the limit $\alpha_j\to \alpha$ for $j=3n,3n\pm1$, we have
\bea
&&
\lim \tau_{n,0}(t)
\non
\\
&&
=
\frac{(-1)^n}{27y_{0}(\alpha)^5}{\rm e}^{-6\sum_{l=1}^\infty\alpha^l t_{3l}}
\sum_{0\leq i<j\leq 2, 0\leq k\leq 2}\omega^{i+k+2j}(1-\omega^{i-j})
\non
\\
&&
\times
\frac{\partial}{\partial \beta}\Bigl(
C_{i,j,k}(\alpha,\beta)
{\rm e}^{\eta(z_i(\alpha)^{-1})+\eta(z_j(\alpha)^{-1})+\eta(z_k(\beta)^{-1})}
\non
\\
&&
\times
\tau_{n-1,0}(t-[z_i(\alpha)]-[z_j(\alpha)]-[z_k(\beta)])
\left.\Bigr)\right|_{\beta=\alpha}.
\label{main-formula}
\ena
for the constants $C_{i,j,k}(\alpha,\beta)$ in (\ref{expression-2}), where $y_{0}(\alpha)$ is given by (\ref{y0}).
\end{theorem}

\begin{remark} 
In the right hand side of (\ref{main-formula}) the exponential factor, which is characteristic to soliton
solutions, is clearly visible. 
Since it can be shown that $\tau_{0,0}=1$ for the genus zero curve $y^3=x$ which corresponds to
the case $\alpha_1=0$, using repeatedly the formula (\ref{main-theorem}) we obtain the formula 
which contains only exponential functions and their derivetaives with respect to parameters. 
The formulas for them were computed in \cite{N2018-1}
independently of Theorem \ref{main-theorem}, where all constants are explicitly given as functions of
 $\{\alpha_j\}$. Thses solutions are called generalized solitons in \cite{N2018-1}.
\end{remark}

\section{The case of hyperelliptic curves}
In this section, based on the results of \cite{BEN},  we derive the corresponding formula to (\ref{main-formula}) in the case of hyperelliptic curve
$X_g$ defined by
\bea
&&
y^2=\prod_{j=1}^{2g+1}(x-\alpha_j)
\label{h-curve}
\ena
and its degeneration 
\bea
&&
\alpha_{2g+1}, \alpha_{2g-1}\to \alpha,
\label{h-limit}
\ena
where $\alpha\neq \alpha_j$ for $1\leq j\leq 2g-2$.
The curve $X_g$ has the unique point over $x=\infty$ which we also denote by $\infty$. We take the local coordinate $z$ around $\infty$ such that
\bea
&&
x=z^{-2}, \qquad
y=z^{-2g-1}F_g(z),
\quad
F_g(z)=\left(\prod_{j=1}^{2g+1}(1-\alpha_i z^2)\right)^{1/2}.
\label{h-coord}
\ena

Let
\bea
&&
\mu^{(g)}=(g,g-1,...,1)
\non
\ena
be the partition and $\tilde{\xi}_g$ a frame of $z^gH^0(X_g,{\cal O}(\ast\infty))$ such that the
corresponding tau function has the expansion of the form 
\bea
&&
\tau(t;\tilde{\xi})=s_{\mu^{(g)}}(t)+\rm{h.w.t.}
\label{h-expansion-1}
\ena
Fix one of the square root $\alpha^{-1/2}$ and define $y_0$ by
\bea
&&
y_0=\alpha^{g-1/2}F_{g-1}(\alpha^{-1/2}).
\label{h-y0}
\ena
Then $(\alpha,y_0)$ is a point of $X_{g-1}$. Set
\bea
&&
p_{\pm}=(\alpha, \pm y_0).
\label{h-ppm}
\ena
Then the values of the local coordinates of $p_\pm$ are
\bea
&&
z(p_\pm)=\pm \alpha^{-1/2}.
\non
\ena

Let $\xi_{g-1}(p_\pm)$ be a frame of $\displaystyle z^{g-2}H^0\left(X_{g-1},{\cal O}(p_\pm+\ast\infty)\right)$
such that their tau functions have the following expansions
\bea
&&
\tau(t;\xi_{g-1}(p_\pm))=s_{\mu^{(g-2)}}(t)+\rm{h.w.t.}
\label{h-expansion-2}
\ena

The following theorem is proved in \cite{BEN} in a similar way to Theorem \ref{relation-1}.

\begin{theorem}\rm{\cite{BEN}}\label{h-theorem1}
The following relation holds.
\bea
&&
\lim \tau(t;\tilde{\xi})
\non
\\
&&
=(-1)^{g-1}(2y_0)^{-1}{\rm e}^{-2\sum_{l=1}^\infty \alpha^l t_{2l}}
\left(\tau(t;\xi_{g-1}(p_+))-\tau(t;\xi_{g-1}(p_-))\right),
\label{h-relation-1}
\ena
where $\lim$ in the left hand side means the limit taking $\alpha_{2g+1}$, $\alpha_{2g}$ to $\alpha$.
\end{theorem}

Let $\tau_{g,0}(t)$ denote the function defined by the right hand side of (\ref{0-point-space}) for $X_g$.

\begin{lemma}\label{h-tau-relations}
(i) $\tau(t;\tilde{\xi}_g)=\tau_{g,0}(t).$
\vskip2mm
\noindent
(ii) For  some constant $C_\epsilon(\alpha)$
\bea
&&
\tau(t;\xi_{g-1}(p_\epsilon))=C_\epsilon(\alpha){\rm e}^{\sum_{l=1}^\infty (\epsilon \alpha^{-1/2})^{-l}t_l}
\tau_{g-1,0}(t-[\epsilon\alpha^{-1/2}]),
\quad \epsilon=\pm.
\non
\ena
\end{lemma}
\par
\noindent
{\it Proof.} (i) Both $\tau(t;\tilde{\xi}_g)$ and $\tau_{g,0}(t)$ are tau functions corresponding to 
$z^gH^0(X_g,{\cal O}(\ast\infty))$. By comparing the expansions (\ref{tau-expansion-1}) and
 (\ref{h-expansion-1})
 we get the result.

\noindent
(ii) Since the right hand side and the left hand side without $C_\epsilon(\alpha)$ of the equation 
in the assertion are the tau functions corresponding to 
$z^{g-2}H^0(X_{g-1},{\cal O}(p_\epsilon+\ast \infty))$ by the definition of $\xi_{g-1}(p_\epsilon)$ and 
Theorem \ref{m-point-space}, the assertion follows. 
\qed

This lemma is proved in \cite{BEN} in a different form. The explicit form of the constant $C_\epsilon(\alpha)$ 
can be extracted from there. Let us give the formula. 

Let $m^{(g)}=\left[\frac{g+1}{2}\right]$. Define the sequence $A^{(g)}$ and $s^{(g)}\in \{\pm1\}$ by
\bea
A^{(g)}&=&(a_1^{(g)},...,a_{m^{(g)}}^{(g)})=(2g-1,2g-5,2g-9,...),
\non
\\
s^{(g)}&=&(-1)^{(g-1)m^{(g)}}.
\non
\ena

\begin{example}
$A^{(1)}=(1)$, $A^{(2)}=(3)$, $A^{(3)}=(5,1)$, $A^{(4)}=(7,3)$.
\vskip2mm
$s^{(1)}=1$, $s^{(2)}=-1$, $s^{(3)}=1$, $s^{(4)}=1$.
\end{example}

The following property of $A^{(g)}$ is known \cite{Onishi2005}\cite{NY2012},
\bea
&&
|A^{(g)}|:=\sum_{j=1}^{m^{(g)} } a^{(g)}_j=\frac{1}{2}g(g+1).
\label{Ag-number}
\ena
Denote the sigma function of $X_{g-1}$ by $\sigma^{(g-1)}(u)$.
Set 
\bea
&&
b_i=(a^{(g-2)}_i+1)/2\in \{1,2,...,g-2\}, 
\quad
1\leq i\leq m^{(g-2)},
\non
\ena
and define 
\bea
&&
\sigma^{(g-1)}_{A^{(g-2)}}(u)=
\frac{
\partial^{m^{(g-2)}}
}{
\partial u_{b_1}\cdots \partial u_{b_{m^{(g-2)} } }
}
\sigma^{(g-1)}(u).
\non
\ena
Then, by Theorem 4.14 of \cite{BEN}, we can deduce that
\bea
&&
C_\epsilon(\alpha)=s^{(g-2)}\sigma^{(g-1)}_{A^{(g-2)}}(-\int_\infty^{p_\epsilon} du)^{-1},
\quad
du={}^t(du_1,...,du_g).
\label{expression-const}
\ena

\begin{lemma}\label{relation-const}
The following relation is valid.
\bea
&&
C_-(\alpha)=(-1)^{g-1}C_+(\alpha).
\label{relation2-const}
\ena
\end{lemma}
\vskip2mm
\noindent
{\it Proof.} It is known that the sigma function satisfies the following relation \cite{Onishi2005}\cite{N2010-1}
\bea
\sigma^{(g-1)}(-u)=(-1)^{\frac{1}{2}g(g-1)}\sigma^{(g-1)}(u).
\non
\ena
By differentiating it we get
\bea
&&
\sigma^{(g-1)}_{A^{(g-2)}}(-u)=(-1)^{\frac{1}{2}g(g-1)+m^{(g-2)}}\sigma^{(g-1)}(u).
\label{parity-1}
\ena
We can easily verify that
\bea
&&
\frac{1}{2}g(g-1)+m^{(g-2)}=g-1 \quad{\rm mod.} 2.
\label{parity-2}
\ena
For the hyperelliptic curve $X_{g-1}$ the following relation holds,
\bea
&&
\int_\infty^{p_-} du=-\int_\infty^{p_+} du.
\label{parity-3}
\ena
The assertion of the lemma follows from (\ref{expression-const}),  (\ref{parity-1}), (\ref{parity-2}), (\ref{parity-3}).
\qed

Substituting the equations of  (i), (ii) in Lemma \ref{h-tau-relations}  into (\ref{h-relation-1})  and using 
(\ref{relation2-const}) we get

\begin{theorem}\label{h-main-theorem}
Let $\tau_{g,0}(t)$ be given by the right hand side of (\ref{tau0}) for the hyperelliptic curve $X_g$ 
defined by (\ref{h-curve}). Then in the limit $\alpha_{2g+1}, \alpha_{2g} \to \alpha$ we have
the following formula,
\bea
\lim \tau_{g,0}(t)&=&(-1)^g(2y_0)^{-1}C_+(\alpha){\rm e}^{-2\sum_{l=1}^\infty \alpha^l t_{2l}}
\non
\\
&&
\times
\Bigl(
{\rm e}^{\eta (\alpha^{1/2})} \tau_{g-1,0}(t-[\alpha^{-1/2}])
+
(-1)^g{\rm e}^{\eta (-\alpha^{1/2})} \tau_{g-1,0}(t-[-\alpha^{-1/2}])
\Bigr),
\non
\ena
where $y_0$, $p_\pm$, $C_+(\alpha)$ are given by (\ref{h-y0}), (\ref{h-ppm}), (\ref{expression-const})
respectively.
\end{theorem}

\begin{remark} The tau function $\tau_{g,0}(t)$ gives a solution of the KdV hierarchy (see remark 
\ref{remark-tau0}).
Again it can be shown that $\tau_{0,0}=1$ for the genus zero curve $y^2=x$ which cooresponds to 
$\alpha_1=0$. Using the formula repeatedly we get the well known soliton solution \cite{NMPZ}\cite{Hirota}.
For $\alpha_1\neq 0$ we can show that $\displaystyle \tau_{0,0}(t)={\rm e}^{L(t)+Q(t)}$, where $L(t)$ and $Q(t)$ are certain linear and quadratic functions of $t$.

\end{remark}

\vskip10mm
\noindent
{\bf \large Acknowledegements}
\vskip2mm
\noindent
I would like to thank Samuel Grushevsky for discussions and helpful comments 
on the degenerations of $Z_N$ curves and associated theta functions,
  and Julia Bernatska and Victor Enolski\footnote{now deceased} for former collaborations.
This work was supported by JSPS KAKENHI Grant Number JP19K03528.


\begin{thebibliography}{99}



\bibitem{AN}
T. Ayano and A. Nakayashiki, 
On Addition Formulae for Sigma Functions of Telescopic Curves, 
{\it Symmetry, Integrability and Geometry: Methods and Applications (SIGMA)} {\bf 9} (2013), 046,
14 pages.


\bibitem{BL}
J. Bernatska and D. Leykin,
On degenerate sigma-functions in genus two,
{\it Glasgow Math. J.} {\bf 61-1} (2019), 169-193,

\bibitem{BEN} J. Bernatska, V. Enolski and A. Nakayashiki, 
Sato Grassmannian and degenerate sigma function,
arXiv:1810.01224


\bibitem{BEL1997}
V. M.~ Buchstaber, V. Z. Enolski and D.~V.~ Leykin,
Kleinian functions, hyperelliptic Jacobians and
applications, in {\it Reviews in Math. and Math. Phys.}
{\bf 10}, No.2, Gordon and Breach, London, 1997, 1-125.

\bibitem{BEL1999}
V. M. Buchstaber, V. Z.  Enolski and D. V. Leykin,
Rational analogue of Abelian functions,  {\it Funct. Anal. Appl.}
{\bf 33} (1999), 83-94

\bibitem{BEL2012}
V.~M.~ Buchstaber, V.~Z.~ Enolski and D.~V.~Leykin,
Multi-Dimensional Sigma-
Functions, arXiv:1208.0990, 2012


\bibitem{DJKM}
E. Date, M. Kashiwara, M. Jimbo, and T. Miwa,
Transformation groups for soliton equations,
in ``Nonlinear Integrable Systems --- Classical Theory and Quantum Theory'',
M.~Jimbo and T.~Miwa (eds.), World Sci., Singapore, 1983,
pp.39--119.

\bibitem{FK}
H.M. Farkas and I. Kra,
{\it Riemann Surfaces},
Second Edition, Springer-Verlag, 1992.


\bibitem{Fay73}
J. Fay, Theta-Functions on Riemann Surfaces, 
Springer Lecture Notes in Mathematics {\bf 352}, 1973.

\bibitem{Hirota} R. Hirota, The direct method in soliton theory, Cambridge University Press,
2004.



\bibitem{KNTY}
N. Kawamoto, Y. Namikawa, A. Tsuchiya and Y. Yamada,
Geometric realization of conformal field theory on Riemann 
surafces, {\it Comm. Math. Phys.} {\bf 116} (1988) 247-308.



\bibitem{Kodama2017}
Y. Kodama, 
KP solitons and the Grassmannians, 
Springer, 2017.

\bibitem{Kr1977}
I.M. Krichever, Methods of algebraic geometry in the theory of nonlinear equations, 
{\it Russ. Math. Surv.}, {\bf Vol. 32} (1977), 185-213


\bibitem{Mac1995}
I. G. Macdonald,
{\it Symmetric Functions and Hall Polynomials, second edition},
Oxford University Press, 1995.


\bibitem{Mul1994} M. Mulase, Algebraic theory of the KP equations, in {\it Perspectives in Math. Phys.},
R.Penner and S.T.Yau (eds.), International Press Company, (1994), 157-223.


\bibitem{N2010-1}
A. Nakayashiki, 
On algebraic expressions of sigma functions for $(n, s)$ curves, 
{\it Asian J. Math.} {\bf 14} (2010), 175–211.

\bibitem{N2010-2}
A. Nakayashiki,
Sigma function as a tau function,
{\it  Int. Math. Res. Not. IMRN} {\bf 2010-3} (2010), 373-394.


\bibitem{N2016}
A. Nakayashiki,
Tau function approach to theta functions, 
{\it Int. Math. Res. Not. IMRN} {\bf 2016-17} (2016), 5202--5248.


\bibitem{N2018-1}
A. Nakayashiki, Degeneration of trigonal curves and solutions of the KP-hierarchy, 
{\it Nonlinearity} {\bf 31} (2018), 3567-3590.


\bibitem{N2018-2}
A. Nakayashiki, On reducible degeneration of hyperelliptic curves and soliton solutions,
{\it SIGMA} {\bf 15} (2019), 009, 18 pages, arXiv:1808.06748.

\bibitem{NY2012}
A. Nakayashiki and K. Yori, Derivatives of Schur, tau and sigma functions, 
on Abel-Jacobi images, in Symmetries, Integrable Systems and Representations,
 K.Iohara, S. Morier-Genoud, B. Remy (eds.), Springer, 2012, 429-462.


\bibitem{NMPZ} S.Novikov, S.V. Manakov, L.P. Pitaevskii and V.E. Zakharov,
Theory of solitons, Consultants Bureau, New York, 1984.


\bibitem{Onishi2005}
Y. \^Onishi, Determinant expressions for hyperelliptic functions,
with an Appendix by Shigeki Matsutani: Connection of the formula of Cantor and Brioschi-Kiepert type,
{\it Proc. Edinburgh Math. Soc.} {\bf 48} (2005), 705-742.

\bibitem{SN}
M. Sato and M. Noumi, Soliton Equations and Universal Grassmann Manifold, 
Sophia University, Tokyo, Mathematical Lecture Note 18, (1984) (in Japanese).


\bibitem{SS}
M.~Sato and Y.~Sato,
Soliton equations as dynamical systems on infinite dimensional Grassmann manifold, 
in ``Nolinear Partial Differential Equations in Applied Sciences'',
P.~D.~Lax, H.~Fujita and G.~Strang (eds.), 
North-Holland, Amsterdam, and Kinokuniya, Tokyo, 1982, pp.259--271.

\end{thebibliography}
\end{document}